\documentclass[conference]{IEEEtran}
\IEEEoverridecommandlockouts
\usepackage{cite}
\usepackage{amsmath,amssymb,amsfonts}
\usepackage{algorithmic}
\usepackage{graphicx}
\usepackage{textcomp}
\def\BibTeX{{\rm B\kern-.05em{\sc i\kern-.025em b}\kern-.08em
    T\kern-.1667em\lower.7ex\hbox{E}\kern-.125emX}}
\begin{document}

\title{The PLATO Payload and Data Processing System SpaceWire network\\
\thanks{This work has been founded by ASI, the Italian Space Agency.}
}

\author{\IEEEauthorblockN{Mauro Focardi}
\IEEEauthorblockA{\textit{INAF-OAA} \\
\textit{Arcetri Astrophysical Observatory}\\
Firenze, Italy \\
mauro.focardi@inaf.it}
\and
\IEEEauthorblockN{Rosario Cosentino}
\IEEEauthorblockA{\textit{INAF-FGG} \\
\textit{Galileo Galilei Foundation}\\
La Palma - Canary Islands, Spain \\
cosentino@tng.iac.es}
\and
\IEEEauthorblockN{Stefano Pezzuto}
\IEEEauthorblockA{\textit{INAF-IAPS} \\
\textit{Inst. of Space Astrophysics and Planetology}\\
Roma, Italy\\
stefano.pezzuto@inaf.it}
\and
\IEEEauthorblockN{David Biondi}
\IEEEauthorblockA{\textit{INAF-IAPS} \\
\textit{Inst. of Space Astrophysics and Planetology}\\
Roma, Italy\\
david.biondi@inaf.it}
\and
\IEEEauthorblockN{Giovanni Giusi}
\IEEEauthorblockA{\textit{INAF-IAPS} \\
\textit{Inst. of Space Astrophysics and Planetology}\\
Roma, Italy\\
giovanni.giusi@inaf.it}
\and 
\IEEEauthorblockN{Luca Serafini}
\IEEEauthorblockA{\textit{KI} \\
\textit{Kayser Italia} \\
Livorno, Italy\\
l.serafini@kayser.it}
\and
\IEEEauthorblockN{Carlo Del Vecchio Blanco}
\IEEEauthorblockA{\textit{KI} \\
\textit{Kayser Italia} \\
Livorno, Italy \\
c.delvecchioblanco@kayser.it}
\and
\IEEEauthorblockN{Donatella Vangelista}
\IEEEauthorblockA{\textit{KI} \\
\textit{Kayser Italia} \\
Livorno, Italy \\
d.vangelista@kayser.it}
\and
\IEEEauthorblockN{Matteo Rotundo}
\IEEEauthorblockA{\textit{Department of Information Engineering} \\
\textit{University of Pisa} \\
Pisa, Italy \\
rotundo.matteo@gmail.com}
\and
\IEEEauthorblockN{Luca Fanucci}
\IEEEauthorblockA{\textit{Department of Information Engineering} \\
\textit{University of Pisa} \\
Pisa, Italy \\
luca.fanucci@unipi.it}
\and
\IEEEauthorblockN{Daniele Davalle}
\IEEEauthorblockA{\textit{IngeniArs} \\
\textit{IngeniArs S.r.l} \\
Pisa, Italy \\
daniele.davalle@ingeniars.com}

\and
\IEEEauthorblockN{and the PLATO DPS Team}
\IEEEauthorblockA{\textit{coordinated by DLR} \\
\textit{German Aerospace Center} \\
Berlin, Germany\\
}
}

\maketitle

%----------
\begin{abstract}
PLATO\cite{b1} has been selected and adopted by ESA as the third medium-class Mission (M3) of the Cosmic Vision Program, to be launched in 2026 with a Soyuz-Fregat rocket from the French Guiana. Its Payload (P/L) is based on a suite of 26 telescopes and cameras in order to discover and characterise, thanks to ultra-high accurate photometry and the transits method, new exoplanets down to the range of Earth analogues. Each camera is composed of 4 CCDs working in full-frame or frame-transfer mode. 24 cameras out of 26 host 4510 by 4510 pixels CCDs, operated in full-frame mode with a pixel depth of 16 bits and a cadence of 25 s.

Given the huge data volume to be managed, the PLATO P/L relies on an efficient Data Processing System (DPS) whose Units perform images windowing, cropping and compression. Each camera and DPS Unit is connected to a fast SpaceWire (SpW) network running at 100 MHz and interfaced to the satellite On-Board Computer (OBC) by means of an Instrument Control Unit (ICU), performing data collection and compression.

\end{abstract}


%----------
\section{Introduction}
The PLATO  Payload, based on two groups of 24 and 2 telescopes, covering ~2250 square degrees per pointing, is conceived for the study of planetary systems formation and evolution and to answer fundamental questions concerning the existence of other planetary systems like our own, including the presence of new worlds in the habitable zone of Sun-like stars.
The 24 telescopes, observing faint stars, and the 2 telescopes, observing bright targets, named respectively \emph{normal} and \emph{fast} telescopes, are operated by their front-end electronics (FEEs, each controlling 4 CCDs as Focal Plane Array -FPA- detectors) and will provide the capability to attain a large photometric visible magnitude range, from ~4 to ~16.
Focusing on a subset of brighter targets (m\textsubscript{v} 4-11), the PLATO Payload will detect and characterise planets down to the Earth size by means of photometric transits and thanks to the determination of their masses through ground-based radial velocity follow-up campaigns.

Given the brightness of this subset of samples, PLATO will extensively perform asteroseismology on these targets to retrieve highly accurate stellar parameters such as masses, radii and ages allowing for a precise characterisation of planetary bulk parameters.

The main scientific requirement to detect and characterise a large number of terrestrial planets around bright stars plays a key role in defining the PLATO observing strategy and its own Payload.

The current baseline observing plan for the 4-years nominal science operations consists of long-duration observations of two sky fields lasting two years each. An alternative scenario is for operations split into a long-duration pointing lasting three years and a one-year step-and-stare phase with several pointings.
Long pointings will guarantee the detection of planets down to the habitable zone of solar-like stars with a first basic assessment of the main characteristics of their atmospheres, opening the way to future space missions (e.g. ARIEL) designed to perform spectroscopy on these targets.

\begin{figure}[htbp]
\centerline{\includegraphics[height=5cm]{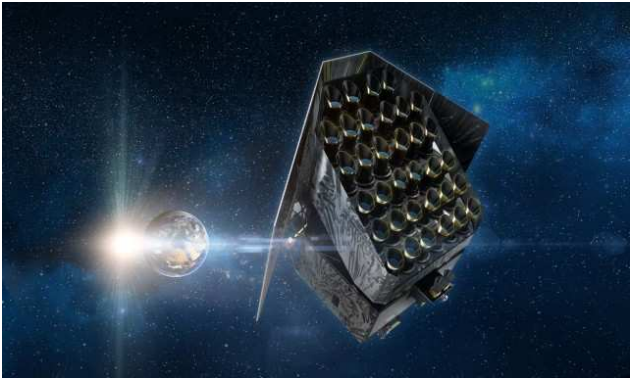}}
\caption{The PLATO satellite with its unique Payload composed by 26 small telescopes in an artistic vision.}
\label{PLATO_DPS}
\end{figure}

The Payload Data Processing System (DPS) is composed of 2 sets of 6 Normal-DPUs (Data Processing Units), each one collecting data from two N-telescopes and 2 Fast-DPUs (one per each Fast camera/telescope) designed to pre-process the downloaded images extracting the photometric signal from the targets and selecting a reduced sample of imagettes (small images of $n$x$n$ pixels) around a subset of stellar samples. The DPUs will also be in charge of the detection and removal of outliers from the photometric signals. 

An Instrument Control Unit (ICU) will collect, thanks to the P/L SpW network, all the scientific data and payload housekeeping (HK) and will compress them before delivery to the Service Module (SVM), as it will act as the main interface between the Payload and the spacecraft (S/C). The ICU will also be in charge of the DPS SpaceWire network management. It is conceived as a full cold-redundant unit implementing a minimal cross-strapping to improve the DPS reliability.

%----------
\section{The PLATO Data Processing System and the P/L SpaceWire network}

The PLATO Payload Data Processing System (DPS, refer to Fig.~\ref{PLATO_DPS}) is made up of the DPUs (Data Processing Units) and the ICU\cite{b2, b3, b4} (Instrument Control Unit), with data routed through a cold redundant SpaceWire network. The ICU is connected to the Service Vehicle Module through four SpW links. These links are independent from the DPS SpaceWire network as the SVM has no direct connection to the other DPS Units, except the Attitude and Orbit Control System (AOCS) interface to two fast DPUs, gathered in one electronic box named FEU (Fast Electronic Unit).

The ICU is responsible for the management of the Payload, the communication with the Service Module and the compression of scientific data before transmitting them as formatted telemetry to the SVM. There are 2 ICU channels, gathered in a same box, which work in cold redundancy and 12 normal DPUs. The SpaceWire routers in the two cold redundant ICU chains can be used in a cross-strapped configuration, e.g. ICU A can use the router from ICU B and vice versa. 

It is the responsibility of the ICU Application SW (ASW) to configure the whole DPS SpaceWire network including all routers in order to achieve the correct data transfer within the Payload. For that purpose, each DPS Units needs a unique SpW node address.
For communication on the SpW network two different protocols are foreseen:

\begin{itemize}
\item the Remote Memory Access Protocol (RMAP), internal to the P/L, and
\item the CCSDS/PUS (Packet Utilisation Standard) packet transfer protocol, internal and external to the P/L.
\end{itemize}

Both RMAP and CCSDS packets are encapsulated in SpW packets. 

A CCSDS packet contains an APID (Application Process ID) and a source or destination identifier. Each DPS unit uses a unique APID for their telecommands (TC) and telemetry (TM) packets. In case of telecommands from the SVM to the PLATO Payload units, the ICU takes the APID to determine the associated SpW node address.

The Payload unit SpaceWire TM packets will provide two logical addresses. One logical address shall be provided for science data packets and one other logical address shall be provided for non-science data packets (housekeeping -HK- packets, acknowledge packets and event packets, etc).

Each N-DPU of the DPS is responsible for processing the data of 2 normal cameras and their processing cadence is nominally 25 sec.

Each F-DPU is responsible for processing the data from only one fast camera. The processing cadence for F-DPUs is 2.5 sec. F-DPUs have a supplementary function: they are responsible for providing angle/attitude error measurements data, as Fine Guidance System (FGS), directly to the SVM AOCS.

\begin{figure}[htbp]
\centerline{\includegraphics[height=7cm]{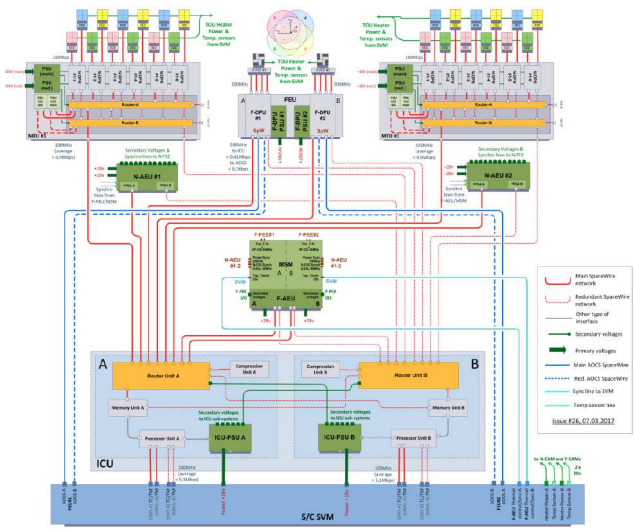}}
\caption{The PLATO Focal Plane Arrays and the P/L electronics Units as connected by the SpaceWire network (red solid lines -nominal- dotted lines -redundant-; blue lines are the SpW links to the AOCS).}
\label{PLATO_DPS}
\end{figure}

After processing, the science data from each N/F-DPU is downloaded to the ICU, which stacks, compresses and then transmits them to the SVM for downloading to Ground. Data from all individual telescopes are transmitted thanks to the on-board transponder to Earth, where final instrumental corrections, such as satellite jitter correction, are performed.

Several photometry algorithms are planned to run on-board, each star being allocated to one of them, depending on its brightness and local level of blending.

Fig.~\ref{PLATO_DPS} represents the PLATO Payload Architecture and it gives an overview of the PLATO Data Processing System with focus on the SpW network (in red) and data routing. All the units of the Payload are connected each other via SpW links and routers (nominal and redundant).

Due to fault tolerance reasons and in order to optimise the resources (mass, volume, harness), the physical implementation of the described architecture foresees to split the 12 N-DPUs in 2 groups of 6 N-DPUs. Each group of 6 N-DPUs is gathered in a box called Main Electronic Unit (MEU).
Also the SpaceWire routers in the two MEUs, as in the ICU, are redundant. They don't implement, as a baseline, an embedded cold-spare functionality, so they shall be protected by an external resistors network, limiting the fault current, and LVDS (Low Voltage Differential Signals) transceivers.

%----------
\section{The Instrument Control Unit, the Data Processing Units and the management of the P/L SpaceWire network}

\begin{figure}[htbp]
\centerline{\includegraphics[height=4.5cm]{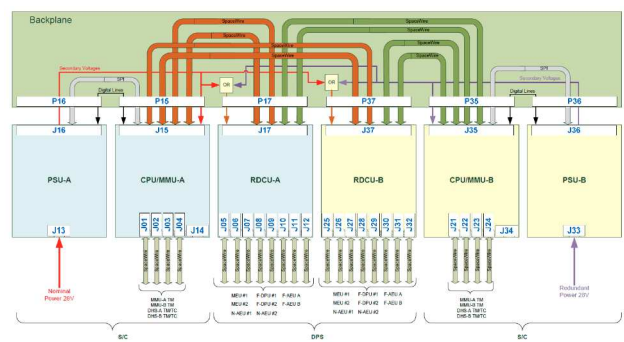}}
\caption{The ICU architecture with its own internal SpaceWire network (orange -Nominal- and green -Redundant- arrows).}
\label{ICU_architecture}
\end{figure}

The ICU (refer to Fig.~\ref{ICU_architecture}) plays a key role in managing the data streams and processing tasks across the entire SpW network, which consists of many nodes and data links. The PLATO SpW network architecture turns out to be crucial due to several design issues:

\begin{itemize}
\item efficient data handling (scientific data, TM and HK);
\item harness complexity;
\item mass and power limitation.
\end{itemize}

Due to the number of data links to be handled, the ICU Router Unit implements an 18-channel SpW router (the GR718B from Cobham-Gaisler) with radiation hardness characteristics (i.e. dedicated to the on-space use). The SpW router board is capable of handling the data streams from the CCD detectors to the ICU in an effective and fast way. The SpW network configuration can be set-up via RMAP commands from the ICU CPU to the router component and the associated control FPGA.

The Instrument Control Unit of PLATO is an integrated box that contains two electronics chains working in cold redundancy.

The main functions of the ICU are the following:

\begin{itemize}
\item Communicate with the Spacecraft (S/C) via SpW links for telecommands and
telemetry;
\item Execute telecommands and forward them to the Data Processing System;
\item Collect scientific and housekeeping data from the PLATO DPS;
\item Compress scientific data and send them to the S/C;
\item Monitor the status of the payload and provide it as telemetry;
\item Perform FDIR (Fault Detection, Isolation and Recovery) tasks and On-board Control Procedures (OBCP) for the effective management of the overall PLATO payload.
\end{itemize}

The S/C provides nominal and redundant +28 V supply voltages to the ICU, which generates
internal secondary voltages for its sub-modules. The ICU is the interface between the
PLATO SVM and the Payload. Considering the high amount of data and the needed data rate, also the ICU subsystems communication is based on an internal SpW network running up-to 100 MHz.

Fig.~\ref{ICU_architecture} shows a block diagram of ICU data and power interfaces as well as its system architecture, as designed by Kayser Italia (ICU industrial Prime Contractor) following the assessment\cite{b5} of IngeniArs Srl, Pisa.

The SpW routing packets between all the PLATO Payload units will be based on SpW logical addressing and the ICU, as the main SpaceWire Network Master, will specify high-level routing changes due to failure and/or redundancy aspects. Accordingly, the PLATO Payload units i.e. MEUs, N-AEUs (Ancillary Electronics Units), F-AEUs, FEU will manage low-level re-configuration of the respective SpW active links. 

In particular, the actual hardware routing path shall be defined by means of routing table within the PLATO ICU. The PLATO Payload units shall update routing paths in response to commands from the ICU.

Also the MEU/DPUs could act, if needed, as SpaceWire Network Master for the upper SpW network level (i.e. with respect to the cameras and their front end electronics). The management of the routing tables will be carried out by them and the PLATO Payload FEEs shall update routing paths in response to commands from the Master Units.

Finally, concerning the P/L and SpW network synchronisation and packets time stamping, the Spacecraft shall provide the means for the PLATO ICU to synchronise its local time to the Spacecraft Elapsed Time (SCET) every second. At every change of seconds of the SCET, the ICU will be provided (through both SpaceWire interfaces from the SVM) with:
 
\begin{itemize}
\item a time tick (SpW Time-code), indicating the precise moment of the change of seconds,
\item a CCSDS/PUS dedicated service providing the value of the SCET at the next time tick.
\end{itemize}

In particular, the CCSDS/PUS Service 9 (Time Management) providing the value of the SCET at the next time tick shall be used solely by the SVM after the ICU bootstrap procedure and successful initialisation.

%----------
\section{Conclusion}
The overall PLATO P/L design relies, as described in this paper, on an efficient cold redundant SpaceWire network. At the present time, thanks to several iterations between ESA, the PLATO Mission Consortium (PMC) and the Spacecraft Prime Contractor (OHB, Bremen), it is under study the possibility to switch-on, when in-flight, both the nominal and redundant sides at the same time in order to manage the remote likelihood to have a double failure on the SpW network and its components. This will be one of the major outcomes of the co-engineering phase between PMC and ESA/Prime, that will end by July, 2018 and will affect the ICU design, under the INAF responsibility.

%----------
\section*{Acknowledgment}

A special acknowledgement to the European Space Agency for the support provided by the PLATO Study Team, to DLR leading the PLATO Mission Consortium and to OHB System AG (Munich, GE), managing and coordinating the development of the overall Payload along with the PMC.

This work has been funded thanks to the Italian Space Agency (ASI) support to the Phases B/C of the Project, as defined within the ASI-INAF agreement n. 2015-019-R.O. "Scientific activity for the PLATO Mission - B/C Phases".

%----------

\end{document}